\documentclass{iau} 
\usepackage{graphicx}

\title[ORLs and CELs in PNe]{The kinematical behavior of ORLs and CELs in PNe with [WC] central star}

\author[Miriam Pe\~na et al.]   
{Miriam Pe\~na$^1$, Francisco Ruiz-Escobedo$^1$, Jackeline Rechy-Garc{\'\i}a$^1$ \and Jorge Garc{\'\i}a-Rojas$^2$
\affiliation{$^1$Instituto de Astronomia, UNAM, M\'exico \\ email: {\tt miriam,fdruiz,jrechy@astro.unam.mx} \\[\affilskip]
$^2$ Instituto de Astrofisica de Canarias, La Laguna, Spain \\email: {\tt jogarcia@iac.es}}
}
\pubyear{}
\volume{323}  
\jname{Planetary Nebulae: Multi-Wavelength Probes of Stellar and Galactic Evolution}
\editors{X.-W. Liu, L. Stanghellini, \& A. Karakas, eds.}
\begin{document}

\maketitle

\begin{abstract}
In Planetary Nebulae (PNe) and HII regions ionic abundances can be derived by using collisionally excited lines (CELs) or recombination lines (ORLs). Such abundances do not coincide for the same ion and usually abundances from ORLs are larger than those from CELs by factors of 2 or larger.  The origin of the discrepancy, known as the Abundance Discrepancy Factor is an open problem in astrophysics of gaseous nebulae. It has been attributed to temperature fluctuations in the plasma, tiny metal-rich inclusions embedded in the H-rich plasma, gas inhomogeneities or other processes. In this work we analyze the kinematical behavior of CELs and ORLs in two PNe ionized by [WC] stars, finding that kinematics of ORLS is incompatible with the kinematics of CELs. In particular the expansion velocities from CELs and ORLs for the same ion are different, indicating that ORLs seem to be produced in zones nearer the central star than CELs. This is in agreement with results found by other authors for individual PNe.
\end{abstract}


\section{Introduction\label{sec:intro}}

PNe are  constituted by  ionized gas surrounding a low-intermediate mass hot  evolved star. The gas  was ejected by the star, at low velocity (of about 10 km s$^{-1}$),  in advanced stages of evolution (Giant or Asymptotic Giant Branch (AGB) stars). Thus the nebula is expanding away from the star with a velocity that depends on the ejection processes, the thermal processes and the interaction of the nebula with the interstellar medium.

The ionized gas emits  collisionally excited lines (CELs) of  ions of heavy element (ions of C, N, O, Ne,  S, and others)  and recombination lines (ORLs) of H, He and ions of heavy elements. Physical conditions (temperature and density) and ionic chemical abundances can be derived from both types of emission lines. Total abundances are obtained by considering all the ions detected in the gas  and  including the not visible ions by correcting the partial abundances with ionization correction factors (ICFs) or other procedures.

It is a well known fact that ionic abundances derived by using CELs or ORLs do not coincide for the same ion and usually the abundances from ORLs are larger than those from CELs by factors of 2 or larger. This occurs in H{\sc ii} regions and PNe as well (see e.g., Garc{\'\i}a-Rojas \& Esteban 2007). In some PNe such a factor, known as the Abundance Discrepancy Factor (ADF), can rise up to 100 or more but the mean value is around 3 (McNabb et al. 2013; Garc{\'\i}a-Rojas et al. 2013). The origin of the ADF is, so far,  an open problem in astrophysics of gaseous nebulae. It has been attributed to temperature fluctuations (e.g., Peimbert 1967, 1971),  tiny metal-rich inclusions embedded in the H-rich plasma (Liu et al. 2006  and references therein), gas inhomogeneities or other processes.
These proposals would imply that CELs and ORLs would be emitted in zones of different temperatures in the nebula, as CELs are highly dependent of the electron temperature, while ORLs are much less dependent of this parameter and they would be  emitted mainly in low temperature zones.

Since some years ago some authors have indicated that the ORLs and CELs emitted by the same ion seem to show different kinematical behavior or different physical distribution in the nebula. In several objects recombination lines of O\,{\sc ii} appears to show lower expansion velocities than [O\,{\sc iii}] collisionally excited lines (e.g.,  Otsuka et al. 2010) or that  the emission of ORLs from  C$^{++}$ and O$^{++}$ 
strongly peak towards the nebular center, see e.g., Liu et al. (2000), Liu et al. (2006) and references therein. 

More recently  Richer et al. (2013) analyzed in detail the spatially and velocity-resolved spectroscopy of ORLs and CELs for the same ion in NGC\,7009, finding that the lines show discrepant kinematics and location. ORLs kinematics is incompatible with the ionization structure given by CELs. These authors found that ORLs arise from a different volume than that giving rise to the forbidden emission from the same ion within this nebula, and  suggest that there is an additional H-deficient plasma component.

Also Garc{\'\i}a-Rojas et al. (2016)  found that the O\,{\sc ii} $\lambda$4649+50 and the [O{\sc iii}]$\lambda$5007 have different spacial distribution in NGC\,6778. These authors claim  that these differences are consistent with the presence of a H-poor gas where ORLs are being emitted. 

We have analyzed the kinematics of ORLs and CELs for some PNe, for which we dispose of very high-resolution spectra, where the lines are well resolved. Most of spectra were obtained  with the Magellan Inamori Kyocera Echelle spectrograph (MIKE, \cite[Berstein et al. 2003]{berstein:03}), attached to the 6.5-m Magellan telescope Clay, at Las Campanas Observatory, Chile.  Most of these data have been already analyzed in order to derive physical conditions and chemical abundances from ORLs and CELs and to determine the ADFs for the objects (Garc{\'\i}a-Rojas et al. 2012; Garc{\'\i}a-Rojas et al. 2013). In these references the procedures for data reduction can be found. Data for other objects have been collected from the literature. The total sample includes PNe ionized by a [WC] central star, a {\it wels}, or a normal. The ADFs for the whole sample varies from 1.5 to 5.

In this article we present partial results for  Cn\,1-5 (PN G002.2-09.4) and PC\,14 (PN G336.2-06.9), ionized by a [WC] central  star, as examples of the whole sample, which will be presented elsewhere.

\vskip -0.2cm
\section{Data analysis}

Due to the high spectral resolution of our data (MIKE has R $\sim$ 28000 for a 1$''$ slit width) the lines are well resolved and in several cases they are split  showing blue and red components coming from the front and back zones of the expanding shell. In other cases the lines appear as a single component or show evidence of several overlapped components. Cn\,1-5 and PC\,14 present split lines and  we  measured both components with IRAF\footnote{IRAF is distributed by the National Optical Astronomy Observatories, which is operated by the Association of Universities for Research in Astronomy, Inc., under contract to the National Science Foundation.}  routines by fitting a Gaussian profile to each line. Thus the observed central wavelength and Gaussian Full Width at Half Maximum ($gfwhm$)  were determined for several ORLs and CELs in each object.  The measured line widths were corrected by the instrumental width, considering that  they add in quadrature. Thermal widths were not corrected, therefore line widths include thermal width, turbulence and  physical structure in the nebula. Partial results for  the CELs and ORLs emitted by O$^+$, O$^{++}$, and N$^+$   are presented in Table 1, where column 1 gives  the ion ID. In columns 2, 3 and 4 the observed central wavelength,  line flux, and $gfwhm$ for lines in Cn\,1-5 are presented, and in columns 5, 6 and 7, the same data is given for PC\,14. As usual, CELs are represented with squared parenthesis, and ORLs without parenthesis. CELs and ORLs  from several other ions  (C$^+$, Fe$^+$, Fe$^{++}$, S$^+$, S$^{++}$, Ne$^{++}$, etc.) were also measured.

Expansion velocities (V$_{exp}$), as given by each CEL and each ORL, were calculated as half the difference between the red and blue components of each line.  The final adopted V$_{exp}$ for each ion, is the average of velocities given by its different CELS and ORLs. They are given at the end of Table 1.

\begin{table}
  \begin{center}
  \caption{Central wavelength (\AA), flux (erg cm$^{-2}$ s$^{-1}$) and gfwhm (\AA) for some CELs and ORLs in Cn\,1-5 and PC\,14.  V$_{exp}$ in km s$^{-1}$.}
  \label{tab1}
 {\scriptsize
  \begin{tabular}{|r|c|c|c|c|c|c|c|}
\hline
  & \multicolumn{2}{c}{Cn\,1-5} & &\multicolumn{2}{r}{PC\,14}&\\
\hline
Ion & center & flux & gfwhm$_o$ & center & flux & gfwhm$_o$ \\
\hline 
[O\,{\sc ii}] & 3726.04 & 2.48E-13 & 0.25 & 3725.23 & 4.89E-14 & 0.29 \\
$[$O\,{\sc ii}] & 3726.69 & 1.64E-13 & 0.25 & 3725.90 & 4.74E-14 & 0.30 \\
$[$O\,{\sc ii}] & 3728.79 & 1.24E-13 & 0.24 & 3727.99 & 2.65E-14 & 0.29 \\
$[$O\,{\sc ii}] & 3729.44 & 8.04E-14 & 0.25 & 3728.66 & 2.49E-14 & 0.29 \\
$[$O\,{\sc iii}] & 4958.94 & 1.36E-12 & 0.47 & 4957.89 & 6.49E-13 & 0.50 \\ 
$[$O\,{\sc iii}] & 4959.80 & 1.09E-12 & 0.53 & 4958.63 & 6.69E-13 & 0.39 \\
$[$O\,{\sc iii}] & 5006.88 & 3.85E-12 & 0.46 & 5005.81 & 2.70E-12 & 0.46 \\
$[$O\,{\sc iii}] & 5007.76 & 3.21E-12 & 0.55 & 5006.56 & 2.96E-12 & 0.38 \\
O\,{\sc ii} & 4076.35 & 7.73E-15 & 0.35 & 4075.49 & 1.91E-15 & 0.48 \\
O\,{\sc ii} & 4077.04 & 6.02E-15 & 0.27 & 4076.21 & 8.53E-16 & 0.25 \\
O\,{\sc ii} & 4649.14 & 1.14E-15 & 0.37 & 4648.45 & 7.08E-16 & 0.30 \\
O\,{\sc ii} & 4649.91 & 1.11E-15 & 0.30 & 4649.94 & 3.19E-16 & 0.25 \\
$[$N\,{\sc ii}] & 6548.09 & 7.42E-13 & 0.54 & 6546.65 & 4.27E-14 & 0.61 \\
$[$N\,{\sc ii}] & 6549.21 & 5.59E-13 & 0.61 & 6547.84 & 3.87E-14 & 0.49 \\
$[$N\,{\sc ii}] & 6583.47 & 2.30E-12 & 0.54 & 6582.02 & 1.28E-13 & 0.61 \\
$[$N\,{\sc ii}] & 6584.60 & 1.73E-12 & 0.61 & 6583.21 & 1.15E-13 & 0.48 \\
N\,{\sc i} & 8223.12 & 2.18E-16 & 0.75 &  &  &  \\
N\,{\sc i} & 8224.43 & 1.66E-16 & 0.80 &  &  &  \\
\hline
V$_{exp}$[O\,{\sc ii}]& &26.3$\pm$0.1 & & &  23.9$\pm$0.3 & \\ 
V$_{exp}$[O\,{\sc iii}]& &26.4$\pm$1.2&&&22.3$\pm$0.4 & \\
V$_{exp}$(O\,{\sc ii}) & &25.4$\pm$1.0 &&&20.8$\pm$1.9& \\
V$_{exp}$[N\,{\sc ii}] & &25.7$\pm$0.2 &&& 27.2$\pm$1.2& \\
\hline 
  \end{tabular}
  }
 \end{center}
\end{table} 

In Fig. 1 we present graphs showing the average V$_{exp}$  for each ion given by CELs (in blue) and ORls (in red) as a function of the ionization potential (IP) of the ion which, considering the stratification in the ionization structure usually found in a nebula, in some way represents the distance of the zone where the ion resides, relative to the central star.

\begin{figure}[bt!]
 \includegraphics[width=2.7 in]{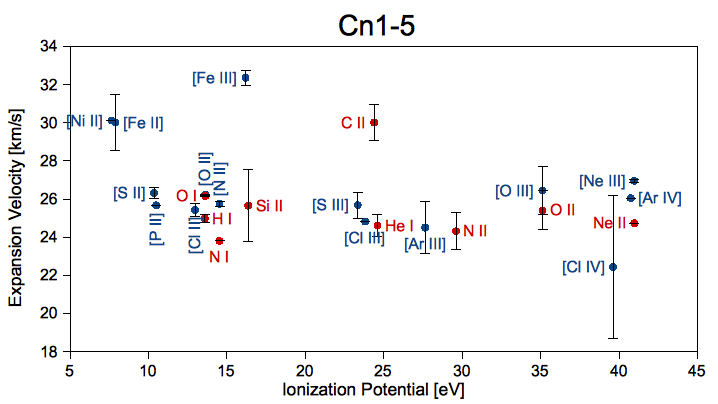}  \includegraphics[width=2.7 in]{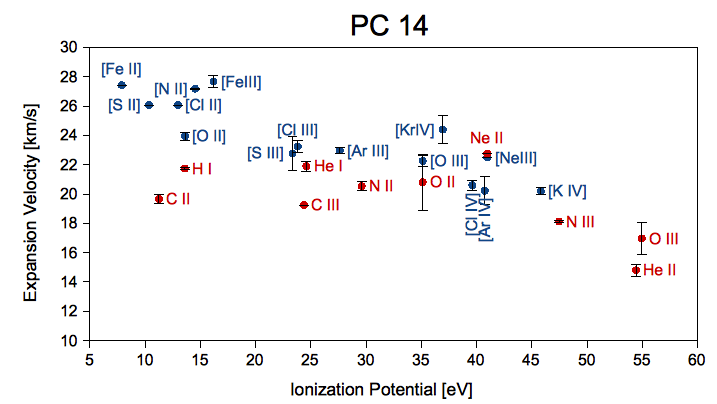} 
 \caption{Expansion velocity vs. ionization potential for different ions as given by CELs (blue) and ORLs (red).}
   \label{fig1}
\end{figure}

\vskip -0.2cm
\section{Results}

Fig. 1  shows evidence of the structure of the velocity field inside the expanding nebulae, in the sense that V$_{exp}$ is lower near the central star and increases outwards. It is expected that low ionization species, which are farther away from the central star, have larger V$_{exp}$ than highly ionized species which lie nearer the star. In Fig. 1 this is particularly true for velocities given by CELs, where a clear gradient is evident for both nebulae. In Cn\,1-5 V$_{exp}$ increases from about 24 km s$^{-1}$ for the highly ionized species ([Cl\,{\sc iv}], [Ar\,{\sc iv}]) to about 30 km s$^{-1}$ for [N\,{\sc ii}] and [Fe\,{\sc ii}] and the same happens with the CELs in PC\,14. For the case of ORLs, there is no gradient or a very flat one. Low- and high-ionization species show about the same V$_{exp}$. 
 In addition, V$_{exp}$ of ORLs are smaller than  V$_{exp}$ of CELs, at the same ionization potential. This is very clear if one considers for instance V$_{exp}$[O\,{\sc iii}] and V$_{exp}$(O\,{\sc ii}) in an object. In both cases presented here V$_{exp}$[O,{\sc iii}] $>$ V$_{exp}$(O\,{\sc ii}), and the same occurs for V$_{exp}$[N\,{\sc ii}] and V$_{exp}$(N\,{\sc i}).

This result recurs in most of the objects under study (Pe\~na et al., in preparation). 

\section{Discussion}
From the results shown above, it seems evident that, as far as CELs and ORLs of the same ion do  not share the same kinematics, these lines are not being emitted in the same zone of the nebula. Considering the expansion velocity  field in the nebula and due to ORLS show lower V$_{exp}$ they seem to be emitted in an inner zone. In this sense we are finding the same results reported by Liu et a. (2000), Richer et al. (2013) and Garc{\'\i}a-Rojas et al. (2016) for NGC\,6153, NGC\,7009 and NGC\,6778 respectively, where ORLs are being mainly emitted in an inner zone than CELs.  And that kinematics of ORLs is incompatible with that of CELs.

In Cn\,1-5 and PC\,14, the zones where ORLs are mainly emitted are not expanding in the same way as the zones where CELs are emitted, and show lower V$_{exp}$. It is tempting to say that ORLs come from H-deficient zones which were ejected by the star after the H-rich gas where CELs are emitted. However at this point this is mainly speculative. Much more analysis is needed, for a larger number of objects.


\vskip -0.2cm

\acknowledgment{ This work has received financial support from grant UNAM-PAPIIT IN109614}.







\begin{thebibliography}{}
\bibitem[\protect\citeauthoryear{Berstein et~al.}{2003}]{berstein:03}
Berstein, R. A.,  Shectman, S. A., Gunnels, S, et al. 2002, \textit{Proc. SPIE} 4841

\bibitem[\protect\citeauthoryear{Garcia-Rojas \& Esteban}{2007}]{garcia:07}
Garc{\'\i}a-Rojas, J., \& Esteban, C. 2007, \textit{ApJ}, 670, 457

\bibitem[\protect\citeauthoryear{Garcia-Rojas  et~al.}{2012}]{garcia:12}
Garc{\'\i}a-Rojas, J., Pe\~na, M., Morisset, C. Mesa-Delgado, A., \& Ruiz, M.T.   2012, \textit{A\&A}, 538, 54

\bibitem[\protect\citeauthoryear{Garcia-Rojas et~al.}{2013}]{garcia:13}
Garc{\'\i}a-Rojas, J., Pe\~na, M., Morisset, C., et al. 2013, \textit{A\&A}, 558, 122

\bibitem[\protect\citeauthoryear{Garcia-Rojas et~al.}{2016}]{garcia:16}
Garc{\'\i}a-Rojas, J., Corradi, R. L. M., Monteiro, H., et al. 2016, \textit{ApJ},  824, 27


\bibitem[\protect\citeauthoryear{McNabb et~al.}{2013}]{mcnabb:13}
McNabb, I. A., et al. 2013, \textit{MNRAS}, 428, 3443

\bibitem[\protect\citeauthoryear{Liu et~al.}{2000}]{liu:00}
Liu, X.-W., Sorey, P. J., Barlow, M. J., et al. 2000,  \textit{MNRAS}, 312, 585

\bibitem[\protect\citeauthoryear{Liu et~al.}{2006}]{liu:06}
Liu, X.-W., Barlow, M. J., Zhang, Y., et al. 2006, \textit{MNRAS}, 368, 1959


\bibitem[\protect\citeauthoryear{Otsuka et~al.}{2010}]{otsuka:10}
Otsuka, M, Tajitsu, A., Hyung, S., \& Izumiura, H. 2010, \textit{ApJ}, 723, 658

\bibitem[\protect\citeauthoryear{Peimbert}{1967}]{peimbert:67}
Peimbert, M. 1967, \textit{ApJ}, 150,825

\bibitem[\protect\citeauthoryear{Peimbert}{1971}]{peimbert:71}
Peimbert, M. 1971, \textit{Bol. Observatorios Tonantzintla y Tacubaya}, 6, 29

\bibitem[\protect\citeauthoryear{Richer et~al.}{2013}]{richer:13}
Richer, M. G., Georgiev, L., Arrieta, A., Torres-Peimbert, S. 2013, \textit{ApJ}, 773, 133

\end{thebibliography}
\end{document}